\documentclass[twocolumn,pra,aps,longbibliography]{revtex4-2}
\usepackage{times}

\usepackage{amsmath,amsfonts,amssymb,graphicx,hyperref,epstopdf,xcolor,tikz,scalerel,natbib,array,gensymb}
\usepackage{mathptmx,textcomp,braket}
\usepackage[T1]{fontenc}
\usepackage[utf8]{inputenc}
\usepackage{multirow}
\usepackage{array}

\definecolor{lime}{HTML}{A6CE39}
\DeclareRobustCommand{\orcidicon}
{
	\begin{tikzpicture} 
	\draw[lime, fill=lime] (0,0) circle [radius=0.15] node[white] {{\fontfamily{qag}\selectfont \tiny ID}};
	\draw[white, fill=white] (-0.0625,0.095) 	circle [radius=0.007];
	\end{tikzpicture}
	\hspace{-2.2mm}
}
\newcommand\orcidID[1]{\href{https://orcid.org/#1}{\orcidicon}}

\newcommand{\Pdd}[2]{\frac{\partial#1}{\partial #2} }
\newcommand{\be}{\begin {equation}}
\newcommand{\ee}{\end {equation}}
\newcommand{\beqa}{\begin {eqnarray}}
\newcommand{\eeqa}{\end {eqnarray}}
\newcommand{\mb}{\mathbf}

\newcommand{\Sch}{Schr\"odinger }
\newcommand{\Exp}[1]{\text{e}^{#1}}
\newcommand{\phink}{\ket{\phi_k^n}}
\newcommand{\Enk}[1]{E_k^#1}
\newcommand{\psik}{\ket{\psi_k(t)}}
\newcommand{\alnk}[1]{\alpha_k^#1(t)}

\hypersetup{colorlinks,citecolor=blue,filecolor=black,linkcolor=blue,urlcolor=blue}

\begin{document}

\title{Temporal control of high-order harmonic cutoffs in periodic crystals}

\author{Nivash R.}
\author{Amol R. Holkundkar\orcidID{0000-0003-3889-0910}}
\email[E-mail: ]{amol@holkundkar.in}
\author{Jayendra N. Bandyopadhyay\orcidID{0000-0002-0825-9370}}

\affiliation{Department of Physics, Birla Institute of Technology and Science - Pilani, Rajasthan,
333031, India}

\date{\today}

\begin{abstract}

A theoretical study on the high-harmonic generation (HHG) in solids by a synthesized driver field is carried out. The vector potential of the driver field is instrumental in determining the Bloch oscillations of the electrons in a periodic crystal, which eventually reflects in the high-order harmonic spectra. To this end, the interaction of the sinc-shaped driver with the periodic crystal is studied. A typical temporal profile of the associated vector potential manifests in extending the harmonic cutoffs compared to the standard sin$^2$ envelopes of similar intensity and duration. It is also observed that the harmonic cutoffs can be controlled in a temporal manner by varying the delay parameter introduced in the proposed sinc-shaped driver, with a well-defined scaling that depends on the energy bands of the periodic crystal under study. Furthermore, it is also observed that the emission time of the cutoff harmonics and even the harmonic cutoff energy can also be controlled by varying the delay parameter systematically. An optimum delay parameter for maximum harmonic cutoff energy and harmonic yield is also deduced.  
\end{abstract}

\maketitle

\section{Introduction}

The last three decades have witnessed tremendous development in the field of high-order harmonic generation (HHG) by atomic and molecular targets and made it possible to have highly tunable extreme-ultraviolet (XUV) pulses in the attosecond regime \cite{RevModPhys.81.163}. The low conversion efficiency in gaseous HHG is one of the major bottlenecks, which is remedied after the advent of the HHG by the  Bloch oscillations in the solids \cite{Ghimire2011,ghimire2012generation, ghimire2019high,wu2015high, you2016anisotropic}. The HHG from the solids is a very lucrative venture  as it promises a compact source of the XUV radiations and attosecond spectroscopy \cite{Luu2015,RevModPhys.90.021002,vampa2017merge,PhysRevA.97.011401,Yue_22_josab}. Various other aspects are also unique for the HHG  obtained  from solids compared to the atomic HHG. For example, the cutoff energy scales linearly with the field amplitude \cite{Liu2017}, multiple plateaus of higher energies \cite{PhysRevA.95.043416},  etc. The transition from the higher conduction bands to the valence band causes the generation of multiple plateaus, which can be understood from the quasi-classical model \cite{Jia_2017}. Some studies are also aimed toward extending the secondary plateaus of higher energies \cite{PhysRevA.94.063403,Ndabashimiye2016,Li_17_opExp}. The inter-band and intra-band dynamics of the  Bloch electrons and the corresponding transitions give rise to rich features in the HHG   from solids. The detailed account of the HHG   from solids can be found in the following seminal review articles \cite{Yue_22_josab,vampa2017merge,RevModPhys.90.021002}. 
Extension of the harmonic cutoffs \cite{PhysRevA.103.053111}, understanding the real-space collision dynamics \cite{PhysRevA.100.043420} in HHG from solids, study of HHG from periodic optical lattices \cite{guan2016high}, time-dependent band population imaging \cite{PhysRevA.95.063419} via HHG in solids, the effect of vacancy defects \cite{PhysRevA.101.013404,Orlando_2022}, HHG in mono-layer and bi-layer Graphene \cite{PhysRevB.103.094308}, and many more interesting studies in the field of the HHG from solids are reported recently. 

The HHG from the solids can be understood from the quasi-classical model \cite{Jia_2017}, wherein some fraction of the electrons near the $k_0=0$ of valence band makes the transition to the conduction band, followed by the intraband dynamics wherein in the momentum space the temporal evolution of the crystal momentum is related to the driving vector potential as $k(t) = k_0 + A(t)$. Once the electron reaches the edge of the Brillouin zone (BZ), then the transition to the next higher conduction band is feasible, resulting in an increased band population of the higher conduction bands. Eventually, an interband transition causes the emission of high-energy photons, bringing back the electron to lower bands. When the electron reaches the edge of BZ, it can also undergo a Bragg's reflection in the same band, and as we mentioned, it can also tunnel to the neighboring conduction band if the bandgap at the BZ boundary is small, this tunneling is referred as the Zener tunneling, and collectively the dynamics are referred as the Bloch-Zener oscillations \cite{Breid_2006}. The vector potential associated with the laser pulse is instrumental in populating the higher-conduction bands in a step-by-step manner. As a consequence, the multiple plateaus in HHG can be understood. The detailed analysis regarding the role of the driver vector potential on the HHG  by periodic crystals is studied in numerous journal articles, e.g., \cite{PhysRevA.97.043413,PhysRevA.95.043416,Du_2017,PhysRevA.94.063403}. 
  
In this work, we study the HHG by an 1D periodic lattice potential driven by the sinc-shaped driver field \cite{Rajpoot_2020,PhysRevA.103.053124,PhysRevResearch.4.033254}. A broad frequency distribution and a single relatively strong field amplitude make the sinc driver's field profile fascinating for generating and controlling the higher-order harmonics \cite{Rajpoot_2020}. The optical frequency combs are routinely used for pulse shaping, which in principle might be a viable source for generating tailor-made field profiles    \cite{Cundiff2003_RMP,Krauss2009,Cundiff2010_NATP,Soto2013_ncomm}. In this work, we relied on the synthesized sinc shaped driver; it is observed that by changing the delay parameter in the driver field, the harmonic cutoff can be controlled temporally in a systematic manner. Furthermore, harmonic cutoff extension is also seen with the sinc shaped driver as compared to the standard sin$^2$ envelopes, and an optimum value for the delay parameter is observed. In principle, the delay parameter used in defining the driver field can be controlled by moving the mirror assembly on an optical bench \cite{Rajpoot_2020}. 

This paper is organized as follows. The theoretical and numerical aspects of the work are presented in Sec. \ref{sec2}, followed by the results and discussion in Sec. \ref{sec3}, and finally conclusion in Sec. \ref{sec4}.

\section{Theory and numerical methods}
\label{sec2} 
 
\begin{figure}[t]
\centering\includegraphics[width=0.9\columnwidth]{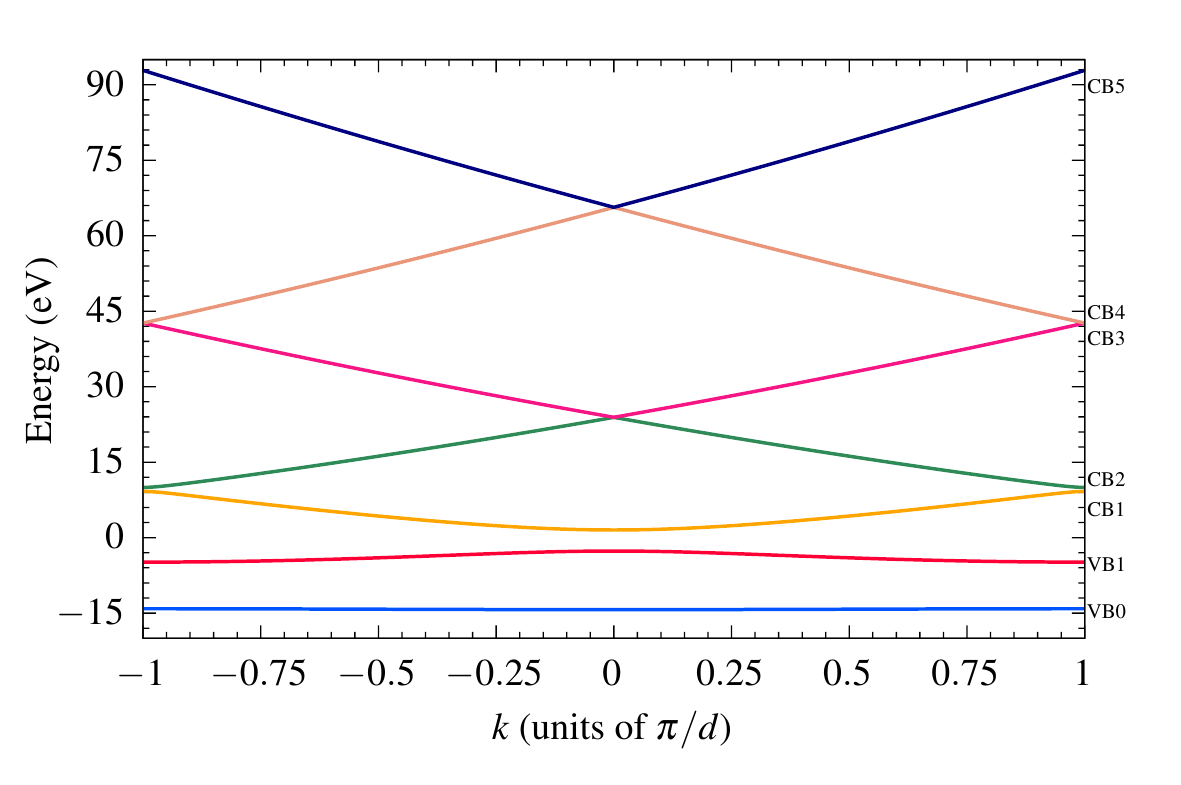}
\caption{The band structure for periodic potential $V(x) = -0.37 [1 + \cos(2\pi x/d)]$ a.u. is calculated for first Brillouin Zone. The bands are labeled as `Valence Band' (VB) and `Conduction Band' (CB). Minimum band gap energy between CB1 and VB1 is $\sim 4.2$ eV.} 
\label{fig1}
\end{figure}

We have studied the laser-solid interaction using the linearly polarized laser propagates along the optical axis of the thin crystal, and the laser polarization direction lies in the crystal plane. The 1D periodic potential used throughout the manuscript is a Mathieu-type potential given as \cite{PhysRev.87.807}: 
\be V(x) = -0.37 [1 + \cos(2\pi x/d)]\ \text{a.u.},\ee 
with $d = 8$ a.u. being the lattice parameter. These types of potential are frequently used in describing optical lattices \cite{Hartmann_2004,PhysRevLett.112.170404}. In this study, we have used $N = 60$ lattice sites, and hence our simulation domain in real space ranges from $-L$ to $+L$ where $L = N d/2$. The typical band structure associated with this periodic potential is calculated and presented in Fig. \ref{fig1}. We have verified that,   for both   sin$^2$ and sinc-shaped driver the results are converged with the inclusion of the 15 bands only \cite{korbman2013quantum}. Though, in Fig. \ref{fig1}, we have only shown   nine bands, wherein VB and CB respectively mean `Valence Band' and `Conduction Band.'

In the following, we present the details of the theoretical and numerical methods  employed in this study. 

\subsection{Laser Pulse}

\begin{figure}[b]
\centering\includegraphics[width=1\columnwidth]{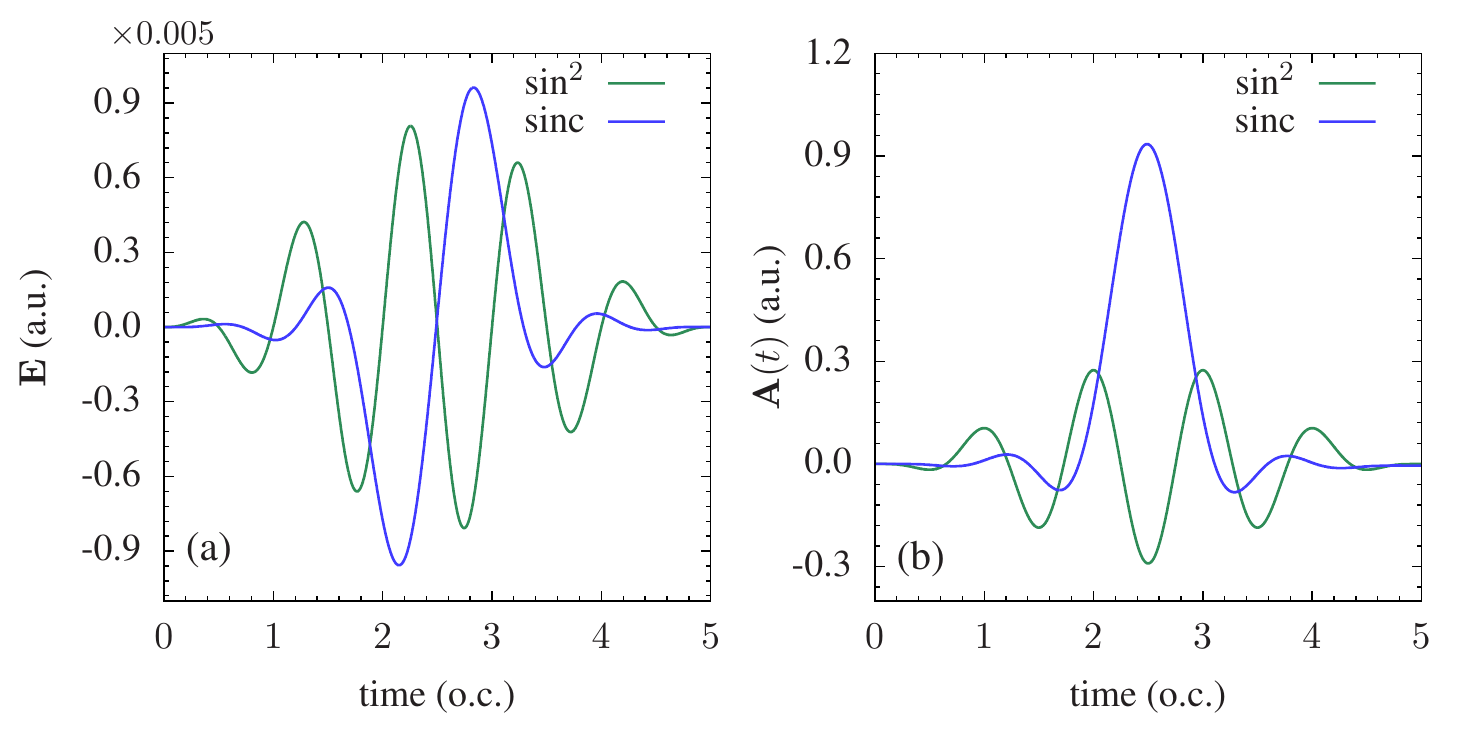}
\caption{The temporal electric field profile of the 3.2 $\mu$m sin$^2$ and sinc laser pulse [$t_d = 0.7\tau$] with a peak intensity of $6\times 10^{11}$ W cm$^{-2}$ are presented (a), along with the respective vector potentials (b). The $y-$axis scaling parameter is also mentioned (a).} 
\label{fig2}
\end{figure}

In the HHG fraternity, it is pervasive to rely on the sin$^2$ field profiles as given by: 
\be \mb{E}_\text{sin2}(t) = F_0 f(t)  \sin(\omega_0 t) \hat{x} \label{field1}\ee
where, $F_0$ is the field amplitude, $f(t) = \sin^2(\pi t/T)$ is the envelope function with pulse duration $T$, and $\omega_0$ is the fundamental driver frequency with one optical cycle being $\tau = 2\pi/\omega_0$. In this work, we use a sinc shape driver pulse, such that the temporal profile of the electric field of the synthesized pulse is written as \cite{Rajpoot_2020}: 
\be
\mb{E}_\text{sinc}(t) = F_0 f(t) \left[ \frac{\sin[\omega_0 (t-t_0-t_d)]}{\omega_0 (t-t_0-t_d) } -  \frac{\sin[\omega_0 (t-t_0)]}{\omega_0 (t-t_0) } \right] \hat{x}, 
\label{field2}
\ee
where $t_d$ is the delay between the pulses and $t_0$ introduces some constant phase. Note that $t_d = 0$ corresponds to the out-of-phase addition of the pulses, which results in $E(t) = 0$. The delay $t_d$ can easily be tuned by varying the mirror assembly on an optical bench \cite{Rajpoot_2020}. It should be noted that the associated vector potential $\mb{A}_\text{sinc}(t) \sim - \int \mb{E}_\text{sinc}(t) dt$ is almost like a sinc function, which can be responsible for vibrant non-linear electron dynamics. The field profiles $\mb{E}_\text{sinc}(t)$ (with $t_d = 0.7\tau$) and $\mb{E}_\text{sin2}(t)$ for some representative parameters are presented in Fig. \ref{fig2}(a). The temporal profile of the respective vector potentials is also shown in Fig. \ref{fig2}(b). We have used a pulse duration $T = 5\tau$ and constant phase factor $t_0 = 2.14\tau$ [so that the pulse peak is temporally at the center] throughout the manuscript. As can be seen from Fig. \ref{fig2}(b), the vector potential for the sinc driver is almost like a sinc function [$\sin(x)/x$], wherein for most of the duration, the $A(t)$ is positive only. 


\subsection{k-space HHG calculation}

  We solve the TDSE numerically in velocity gauge \cite{korbman2013quantum}. The electron wave function can be expanded in Bloch state basis $\phink$ for a particular value of the crystal quasimomentum $k$ and band index $n$. The Bloch states are evaluated by solving the single-electron stationary \Sch equation with field-free Hamiltonian $\hat{H}_\text{o} = \hat{p}^2/2 + V(x)$:
\be \hat{H}_\text{o} \phink = \Enk{n} \phink.\ee
In position basis, the Bloch states can be written as:
\be \braket{x|\phi_k^n} \equiv \phi_k^n(x) = \sum_{\ell=1}^{N_\text{max}} C_{k,\ell}^n\ \Exp{i (k + 2\pi \ell/d) x},\ee
where $N_\text{max} = 15$ are used throughout the work. After the evaluation of the Bloch states, the TDSE can be solved for electronic wavefunction $\psik$ as :
\be i \Pdd{}{t}\psik  = [\hat{H}_\text{o} + \hat{H}_\text{int}] \psik, \label{tdse_sol}\ee
where, $\hat{H}_\text{int} = \mb{A}(t) \cdot \hat{p}$ and $\mb{A}(t)$ is the vector potential associated with the laser pulse under dipole approximation. Furthermore,  
\be \psik = \sum_{n=1}^{N_\text{max}} \alnk{n} \phink, \label{psi_ex}\ee 
where $\alnk{n}$ are the time dependent expansion coefficients. Using Eq. (\ref{psi_ex}) in Eq. (\ref{tdse_sol}), we have coupled differential equations as \cite{korbman2013quantum}:
\be i \Pdd{\alnk{s}}{t}  = \Enk{s} \alnk{s} + A(t) \sum_{u=1}^{N_\text{max}} p_k^{su} \alnk{u}, \label{coeff_sol}.\ee  
Here, $p_k^{su}$ is the matrix element of the momentum operator, which can be calculated as:
\be p_k^{su} = \braket{\phi_k^s|\hat{p}|\phi_k^u} = \sum_{\ell=0}^{N_\text{max}} (k + 2\pi\ell/d) \left(C_{k,\ell}^s\right)^{*} C_{k,\ell}^u.\ee
If we consider the electron initially in the band $q$, then the initial condition for solving Eq. (\ref{coeff_sol}) is $\alpha_k^s(0) = \delta_{qs}$. Finally, the single electron current density for a particular channel $k$ can be calculated as:
\be j_{ks}(t) = - \text{Re}[\braket{\psi_{ks}|\hat{p} + A(t)| \psi_{ks}}]. \label{current}\ee
In Eq. (\ref{current}), the subscript `$s$' denotes that the electron was in the band $s$ before the interaction.
Total current density can be calculated by summing over all the bands and integrating over BZ as:
\be j(t) = \sum_{s\in VB} \int j_{ks}(t) dk\label{current2}.\ee
 In this Bloch state basis based formulation, there is no coupling between different values of $k$, which makes it very straightforward to implement. However, for $k$-space-based calculation of inter and intra-band currents, one must rely on the Houston state basis \cite{wu2015high}. The spectra of the emitted harmonics can be estimated by doing the Fourier transform of the current density and are given as $S(\omega) = \left|\int j(t) \Exp{i \omega t} dt \right|^2$. The harmonic yield $Y$ for the frequency range $\omega_1$ to $\omega_2$ is calculated as: $Y = T^{-1} \int_{\omega_1}^{\omega_2}  S(\omega) d\omega$. 
 
\section{Results and Discussions}
\label{sec3}

Here, we first compare the HHG spectra for the sin$^2$ and sinc-shaped driver, then the delay parameter's effect on the HHG spectra is explored.

\begin{figure}[!t]
\centering\includegraphics[width=1\columnwidth]{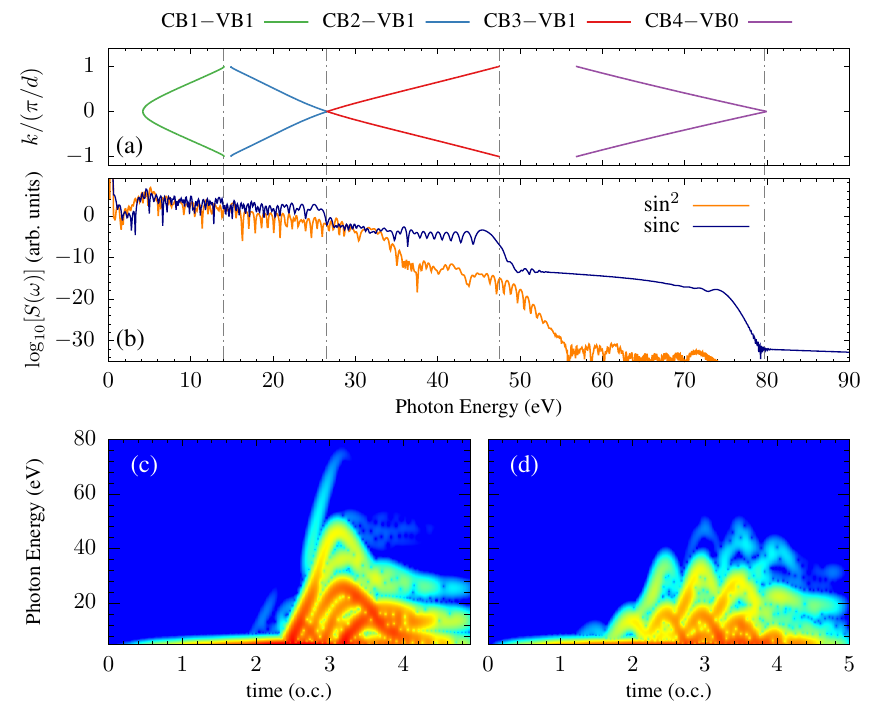}
\caption{The HHG spectra for the sin$^2$ and sinc drivers with the same parameters as in Fig. \ref{fig2}(a) are presented (b) along with the Gabor transform for sinc driver (c) and sin$^2$ driver (d). The respective energy difference of different bands is also illustrated (a) to understand the observed cutoffs.} 
\label{fig3}
\end{figure}

\subsection{Comparing HHG by sinc and sin$^2$ driver}

In Fig. \ref{fig3}, we have compared the harmonic spectra as obtained by the 3.2 $\mu$m laser pulse with peak intensity $6\times 10^{11}$ W/cm$^2$ and having the sin$^2$ envelope [Eq. (\ref{field1})] and sinc shaped field profile with $t_d = 0.7\tau$ [Eq. (\ref{field2})]. It can be seen that the harmonic cutoff in the case of the sinc driver is extended well beyond the one obtained using the sin$^2$ driver. The corresponding energy difference of different bands is also shown in Fig. \ref{fig3}(a), and the multi-plateau structure with the respective harmonic cutoff for sinc shaped driver can easily be mapped on the energy difference between the bands. For the sinc-shaped driver, we see 4 plateaus respectively at $\sim 14$ eV, $\sim 26$ eV, $\sim 47$ eV, and $\sim 79$ eV, which corresponds to the transition from CB1$\rightarrow$VB1, CB2$\rightarrow$VB1, CB3$\rightarrow$VB1, and CB4$\rightarrow$VB0 respectively [Fig. \ref{fig3}(a,b)]. The Gabor transform (time-frequency analysis) for sinc, and sin$^2$ shaped driver are respectively presented in Fig. \ref{fig3}(c) and \ref{fig3}(d). The higher harmonic cutoff for the sinc driver is also seen in the time-frequency analysis. Furthermore, as can be seen from Fig. \ref{fig3}(b), for the sinc driver, the harmonics near the $\sim 47$ eV and $\sim 79$ eV exhibit very smooth modulation. This can be understood from the single trajectory contribution to the harmonic spectra, which is corroborated by Fig. \ref{fig3}(d). In this figure, the higher-energy harmonics emitted only once because the interference from the multiple trajectories is absent, giving very smooth single trajectory contribution. The single trajectory contribution is always sought as a single attosecond pulse can be synthesized using such harmonics instead of the attosecond-pulse train \cite{PhysRevA.84.033414}. The extension of the harmonic cutoff in the case of the sinc driver can be attributed to the vector potential associated with the same [Fig. \ref{fig2}(b)]. The sinc-shaped vector potential does not have a very rapid oscillatory part; as a result, rapid oscillations of the electron in the band are suppressed. The approximate `uni-directional' nature of the vector potential drives the electron in the same direction for most of the pulse duration. As the electron reaches the edge of the BZ, then with Zeneer tunneling, it can climb up the bands  \cite{PhysRevA.97.043413}.  
  
\begin{figure}[t]
\centering\includegraphics[width=1\columnwidth]{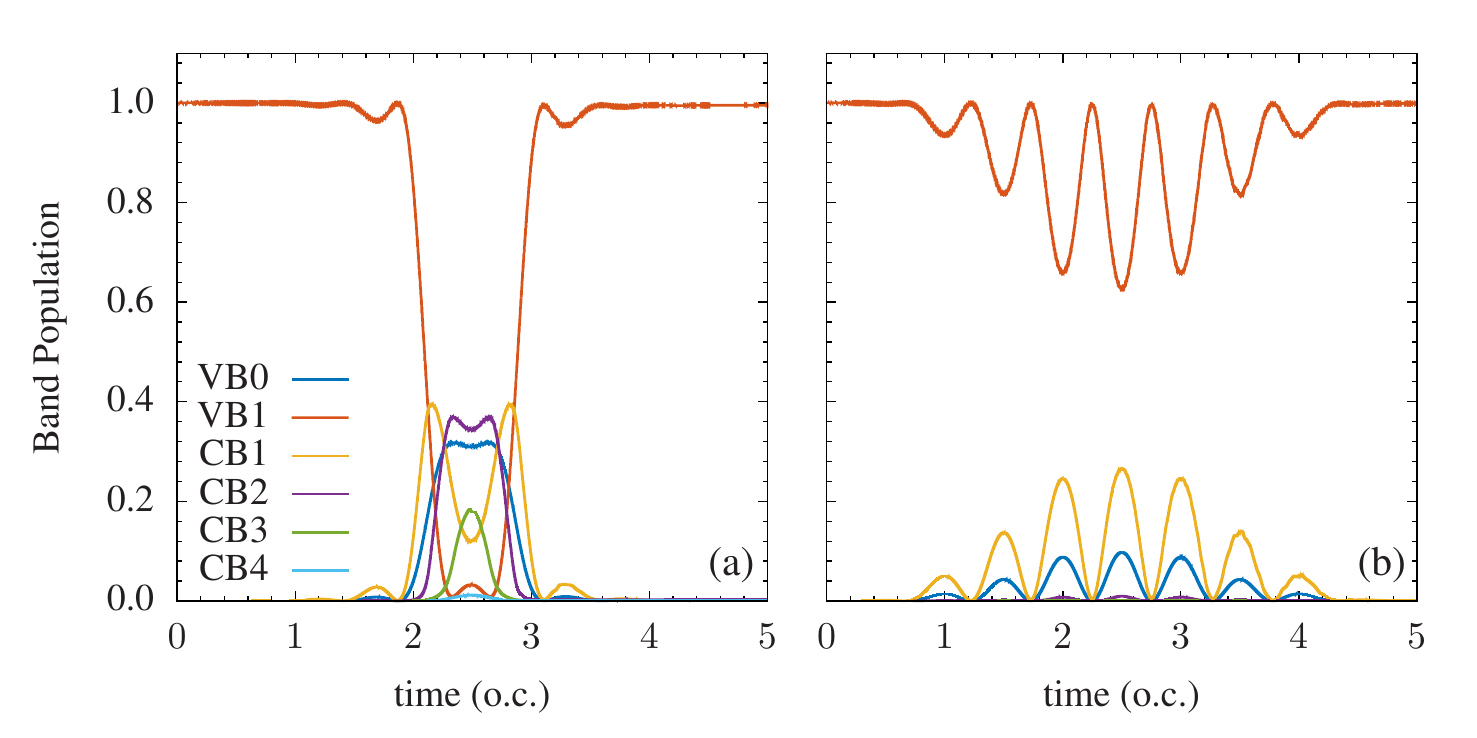}
\caption{Temporal dependence of the band population of various bands [as labeled in Fig. \ref{fig1}] are presented for the sinc (a) and sin$^2$ (a) driver. For both cases, the laser parameters are the same as in Fig. \ref{fig3}.} 
\label{fig4}
\end{figure}

Furthermore, in Fig. \ref{fig4}, the temporal evolution of the band population \cite{chen2019_modphysb} using the sinc and sin$^2$ driver are presented. The time dependent population of the $n^\text{th}$ band is given by $\mathcal{P}_n(t) = \int_{BZ} |\alpha^n_k(t)|^2 dk$ such that $\sum_{n=1}^{N_{max}} \mathcal{P}_n(t) = 1$.
The laser parameters are the same as Fig. \ref{fig3}. Initially, the electron is considered to be in the VB1 in both cases. As can be seen, using the sinc driver, the higher bands are populated, and an appreciable population of even the CB4 band is observed, which in turn results in the emission of the harmonics till $\sim 79$ eV. However, for sin$^2$ pulse, the bands beyond CB2 are hardly populated, resulting in lower cutoff energies. In the following, we study the effect of delay parameter $t_d$ of the sinc driver on the harmonic cutoff and the harmonic yield. 

We observed that the harmonic efficiency beyond $\sim 47$ eV is drastically reduced, which is understandable as the probability of electron tunneling to higher conduction bands in a step-by-step manner reduces. In view of this, we now focus on the HHG spectra till $\sim 50$ eV, which covers the transition from the CB3$\rightarrow$VB1 maximally. 

\subsection{Temporal control of HHG cutoff}

The delay parameter associated with the sinc pulse [Eq. (\ref{field2})] is very instrumental in determining the field profile and so the temporal profile of the vector potential \cite{Rajpoot_2020}. From the experimental perspective, the variation of the delay parameter is equivalent to moving the mirror assembly on an optical bench, which brings us to the question, is it possible to control the band population and so the harmonic cutoff energies and the harmonic yield by simply changing this delay parameter? Given this, we studied the HHG by varying the delay parameter $t_d$ and connected the same to the temporal evolution of the band population.

\begin{figure}[b]
\centering\includegraphics[width=1\columnwidth]{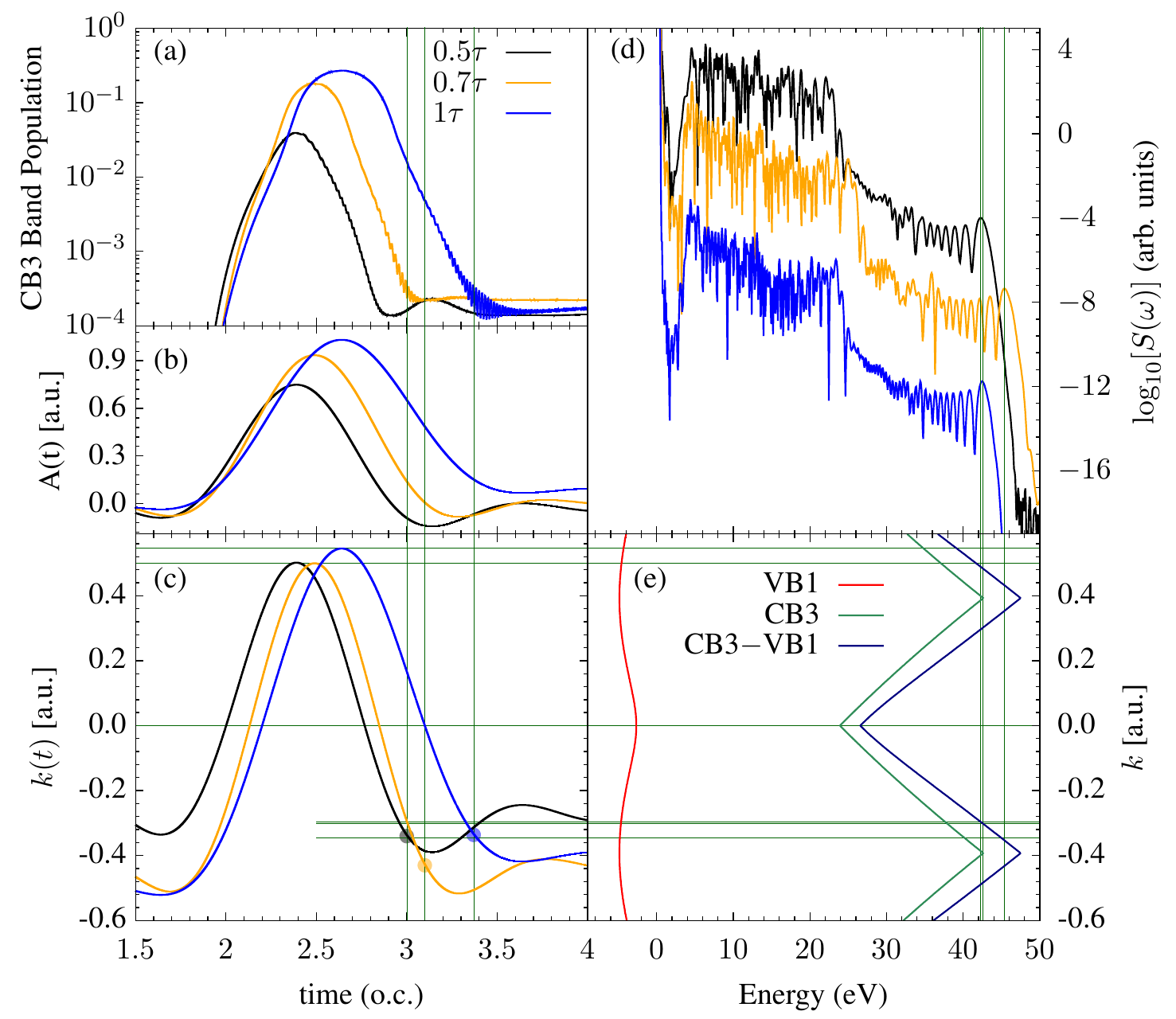}
\caption{Temporal evolution of band population of CB3 (a) along with the vector potential $A(t)$ (b) and crystal momentum $k(t)$ (c), and the HHG spectra (d) is presented for different delay parameter of sinc driver. The bands CB3 and VB1 and their respective energy difference is also shown (e). The HHG spectra for $t_d = 0.7\tau$ and $t_d = 1\tau$ cases are intentionally shifted downward by 4 and 8 units, respectively, for better representation. In (c), the black, orange, and blue circles at $3\tau$, $3.1\tau$, and $3.37\tau$ show the time instants when the maximum harmonic energy is observed for different delay parameters.} 
\label{fig5}
\end{figure}

\begin{figure}[!t]
\centering\includegraphics[width=1\columnwidth]{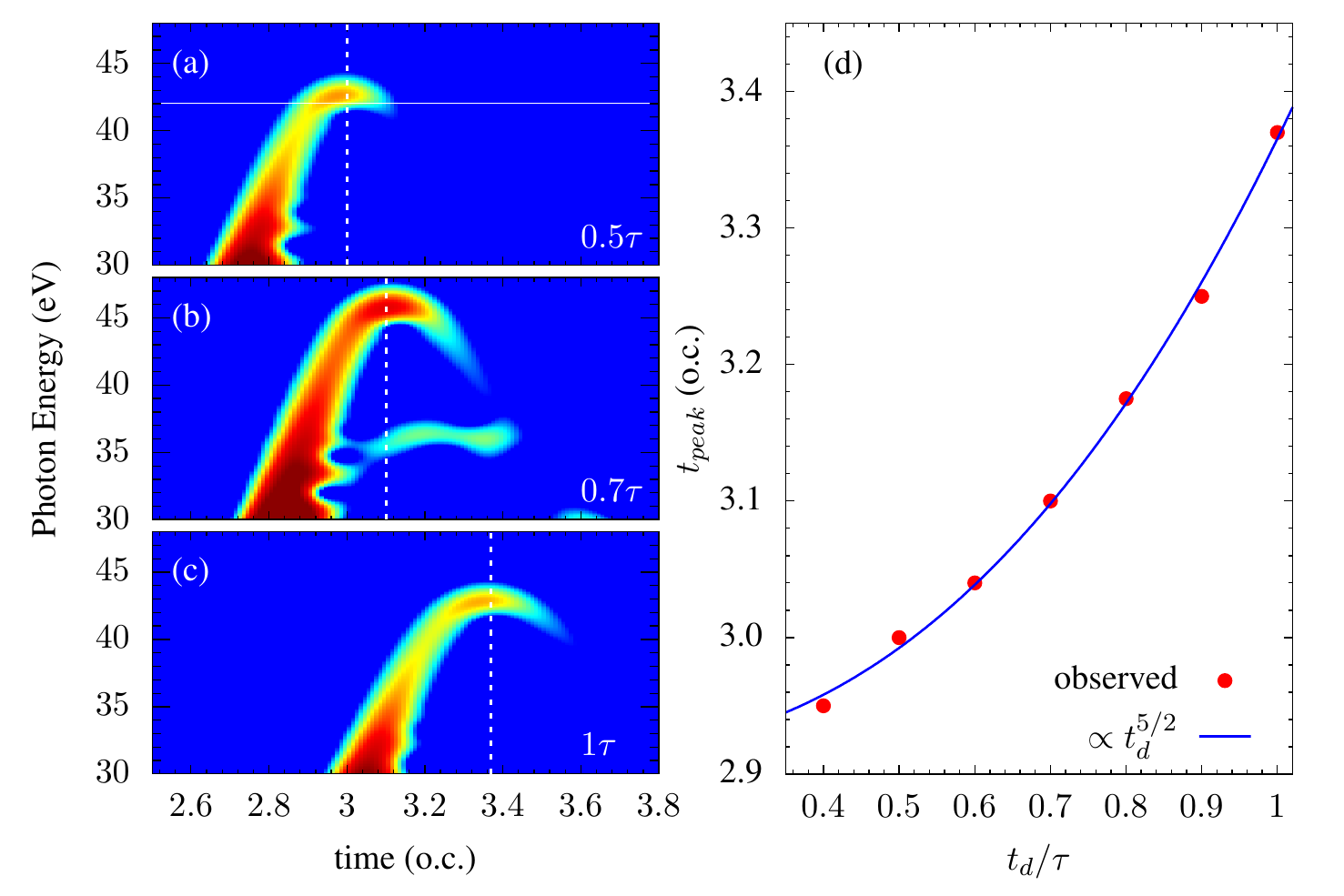}
\caption{The Gabor transform of the HHG spectra for the delay parameters $t_d = 0.5\tau$ (a), $0.7\tau$ (b), and $1\tau$ (c) are presented. All the other laser conditions are the same as in Fig. \ref{fig5}. A vertical dashed white line at $3\tau$ (a), $3.1\tau$ (b), and $3.37\tau$ (c) represents the time instants when the maximum energy harmonics are emitted. The variation of the time instant corresponding to the harmonic cutoff $t_{peak}$ with the delay parameter is presented in (d), wherein the red dots denote the observed values from the simulation and the solid line represents the $\propto t_d^{5/2}$ scaling.} 
\label{fig6}
\end{figure}

\begin{figure}[!t]
\centering\includegraphics[width=1\columnwidth]{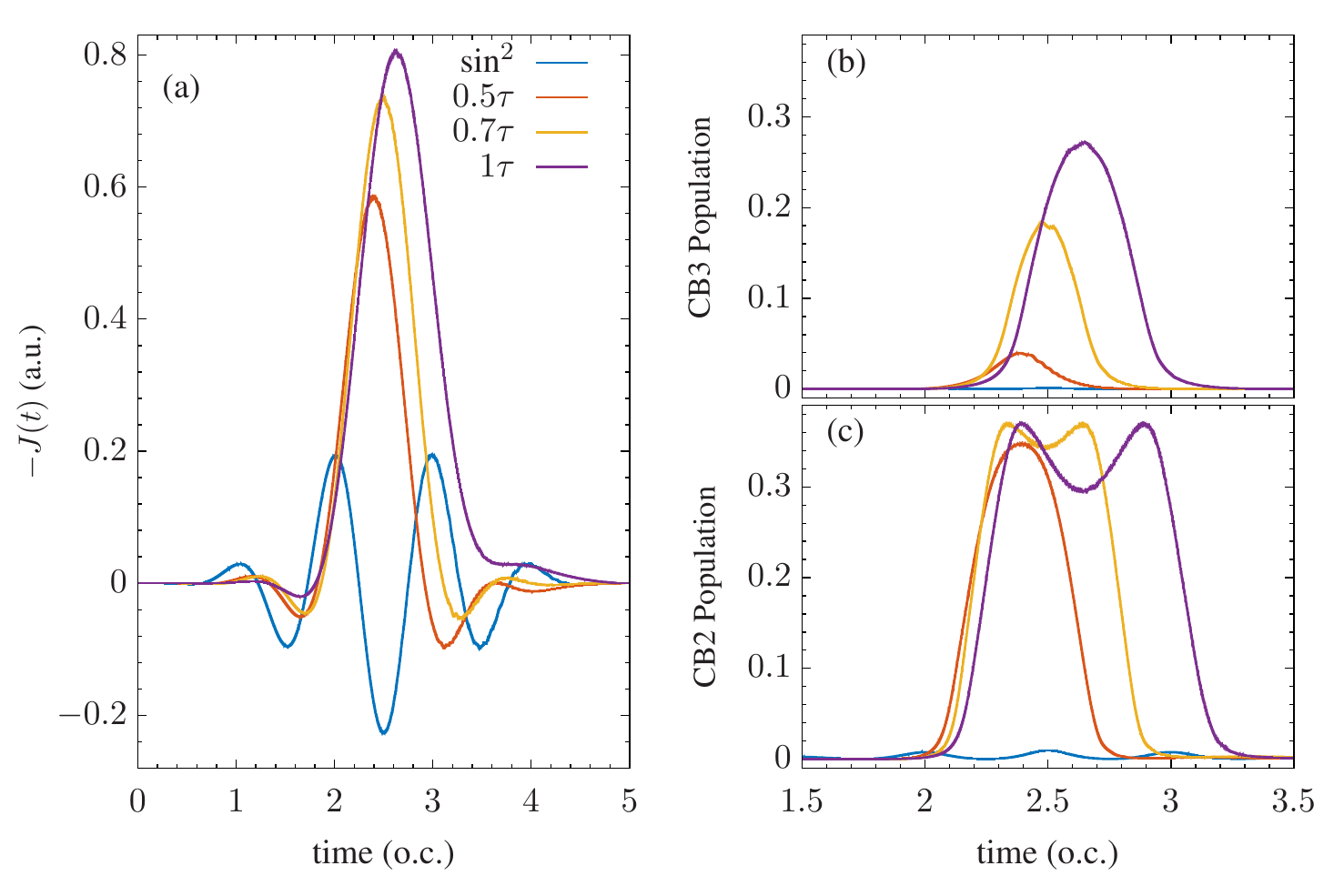}
\caption{Time evolution of the total integrated current [Eq. (\ref{current2})] (a) are presented for sin$^2$ pulse and for the sinc driver with the time delay $0.5\tau$, $0.7\tau$, and $1\tau$. The temporal evolution of the CB3 (b) and CB2 (c) populations is also illustrated using different laser profiles.} 
\label{fig7}
\end{figure}

In Fig. \ref{fig5}(a) we present the temporal evolution of CB3 band-population for $t_d = 0.5\tau, 0.7\tau$ and $1 \tau$ cases. The respective vector potential for these parameters is also shown in Fig. \ref{fig5}(b). The HHG spectra for these three cases are illustrated in Fig. \ref{fig5}(d). Please note that in Fig. \ref{fig5}(d), we have intentionally shifted the HHG spectra for $0.7\tau$ and $1\tau$ for better visual representation. In Fig. \ref{fig5}(e) we have shown the VB1 and CB3 bands in the extended Brillouin Zone [$ |k| > \pi/d$] along with the energy difference between the two [$(E_{CB3} - E_{VB1})$ eV]. As can be seen from Fig. \ref{fig5}(d) and (e) that the harmonic cutoff energies for all three cases can be mapped on the transitions from the CB3 to VB1 band. As in this figure, we are focusing on the harmonic cutoff caused by the transition from the CB3 to VB1, and hence in Fig. \ref{fig5}(c), we present the temporal evolution of the crystal momentum $k(t) = k_0 + A(t)$, where $k_0$ is chosen from the classical trajectories such that for $t_d = 0.5\tau\ ;\ k(2\tau) = 0$, $t_d = 0.7\tau\ ;\ k(2.134\tau) = 0$ and $t_d = 1\tau\ ;\ k(2.2\tau) = 0$. The electron is expected to be in CB3 at $k = 0$ (lowest energy of CB3); however, how and when it will make transition to CB3 is highly unpredictable. These time instances at which the electron is considered to oscillate in CB3 are back-traced by the knowledge of the time-frequency analysis of the emitted harmonics [refer Fig. \ref{fig7}], the classical trajectory calculations, and also noting the appropriate time for the transition from the CB3 to VB1,  resulting in the harmonic cutoff. The time instants at which maximum harmonic energy (cutoff energy) is emitted for  $t_d = 0.5\tau, 0.7\tau,\ \text{and} 1\tau$ are shown in Fig. \ref{fig5}(c) with solid circles at $3\tau, 3.1\tau$ and $3.37\tau$ respectively, which also corroborates the time instants when the CB3 band population is lowest, signifying that after the transition from CB3 to VB1 would seize to exist and hence the abrupt cutoff is observed. In order to further confirm these findings, in Fig. \ref{fig6}(a,b,c), we have presented the Gabor transforms for $t_d = 0.5\tau, 0.7\tau,\ \text{and} 1\tau$ cases. It can be observed that the maximum harmonic energy is emitted at respectively  $3\tau, 3.1\tau$, and $3.37\tau$. The variation of the time instant at which maximum energy is emitted ($t_{peak}$) with the sinc pulse delay parameter $t_d$ is also shown in Fig. \ref{fig6}(d). The red dots in Fig. \ref{fig6}(d) are the observed values from the simulation, and the solid line represents the scaling with the delay parameter, which is found to be $t_{peak} \propto t_d^{5/2}$. This very well-defined non-linear scaling with the delay parameter translates to merely moving the mirror assembly (from an experimental perspective); one can have fine control over the emission time of maximum or cutoff harmonic energy. This particular scaling with the delay parameter $\propto t_d^{5/2}$ is a property of the periodic lattice and the corresponding band-structure. For example, in Appendix-\ref{appen1} we have obtained different scaling for the  same periodic lattice with different lattice constant.   This is so, because the driving of the electron in a band and the follow-up phenomenon of Zener tunneling to higher bands strongly depend on the respective band structure and the minimum band gap near the edge of BZ or at the center of BZ. 
 
Next, in Fig. \ref{fig7}, we compare the temporal evolution of the total current (a), population of the conduction band CB3 (b) and CB2 (c), using sin$^2$ pulse and different delay parameters of the sinc-shaped pulse. It can be seen from Fig. \ref{fig7}(a) that the sinc driver can drive a very strong current in the lattice, closely mimicking the vector potential of the driver. The optical response of the lattice under study can be understood in terms of the polarization, which is defined as  $ P(t) = \int_{-\infty}^t j(t') dt'$ \cite{korbman2013quantum}. As we observe from Fig. \ref{fig7}(a), the time average of the currents in the case of the sin$^2$ driver would be smaller than the sinc driver. The currents caused by the Bloch oscillations using a sinc driver can lead to a very strong polarization response in the solids; mostly because of the same polarity of the current. The strong current in the case of the sinc driver can be understood in terms of the temporal evolution of the CB2 and CB3 band population, contributing the harmonics till $\sim 50$ eV. In Fig. \ref{fig7}(b) and (c), we have compared the temporal evolution of the CB3 and CB2 band population using sin$^2$ and a sinc-shaped driver. The strong vector potential associated with the sinc driver plays a crucial role in populating the higher energy bands by $\sim 5 - 35\%$ or so, while using the sin$^2$ driver with similar peak field amplitude is not observed. 
 
\subsection{HHG cutoff and yield using sinc driver} 

\begin{figure}[!b]
\centering\includegraphics[width=1\columnwidth]{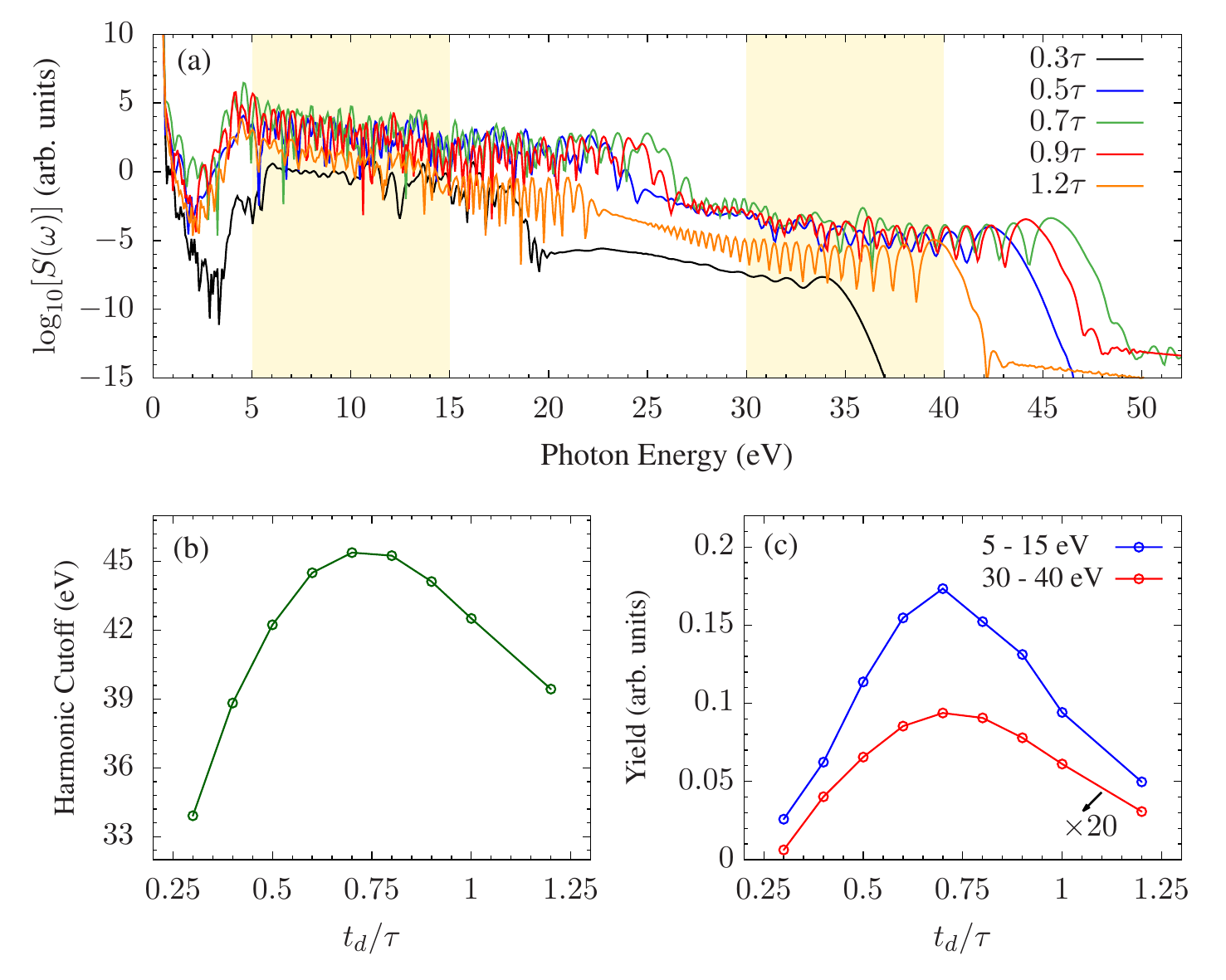}
\caption{Harmonic spectra for different delay parameters are presented (a), and the energy ranges 5 - 15 eV and 30 - 40 eV are highlighted. The variation of the harmonic cutoff energy (b) and harmonic yield (c) with the delay parameter is presented. The harmonic yield in (c) is presented for the two energy ranges highlighted in (a), and the yield for 30 - 40 eV is scaled up by 20 times for better representation.} 
\label{fig8}
\end{figure}

In Fig. \ref{fig8}, we have further explored the effect of the delay parameter on the HHG cutoff and the harmonic yield. We have presented the HHG spectra for different delay parameters of the sinc driver and the variation of the harmonic cutoff energy [Fig. \ref{fig8}(b)] and harmonic yield [Fig. \ref{fig8}(c)] as a function of the delay parameter $t_d$. In Fig. \ref{fig8}(a), two energy ranges are highlighted, for which the harmonic yield is presented in Fig. \ref{fig8}(c). It is observed that there is an optimum delay parameter around $t_d \sim 0.7\tau$ which is responsible for the maximum cutoff energy. The maximum harmonic yield in two energy ranges highlighted in Fig. \ref{fig8}(a) are also presented, and $t_d \sim 0.7\tau$ is observed to be an optimum delay parameter. This optimum delay parameter can be understood from the Fig. \ref{fig5}(b),(c), and (e), wherein it can be observed that the stronger vector potential can drive the electron past the BZ boundary (i.e., the equivalent of the coming from opposite side because of the periodicity of the band-structure), and as a result, the lower harmonic cutoff is expected. We have seen in Fig. \ref{fig5}(b) that the vector potential amplitude for the $t_d = 1\tau$ case is large as compared to the $t_d = 0.7\tau$ case, the corresponding time-dependent crystal momentum $k(t)$ in Fig. \ref{fig5}(c) for $t_d = 1\tau$ case would cause the driving of the electron from $k(2.2\tau) = 0$ to the maximum $k(2.64\tau) = 0.55$, which eventually need to have longer trajectory as compare to the other two cases, to recombine around $3.37\tau$. So in principle, there will be competing mechanisms; the stronger vector potential will try to push the electrons past the edge of the BZ, but in a process, eventually ends up having large dispersion of electronic wavefunction, resulting in lower harmonic yield and the harmonic cutoff (harmonic cutoff would depend on the extent the electron is moved in a band).      

From Fig. \ref{fig8}, we can see that using a sinc driver, we can temporally control the harmonic cutoff. The harmonic cutoff energy can also be tailored by adjusting the delay parameter in a systematic way. The control over harmonic yield in two sample energy ranges also adds to the utility of the sinc drivers. 
 
\section{Summary} 
\label{sec4}

In summary, we have studied the interaction of the sinc-shaped driver with the 1D periodic potential, which is aligned along the polarization direction of the driving laser pulse. The vector potential associated with the proposed field profile [refer Eq. (\ref{field2})] mimics the sinc function, which for most of the pulse duration does not change the polarity. This typical characteristic of the vector potential drive the electron resulting in a very efficient Zener tunneling to higher conduction bands at the minimum-band gap of the neighboring bands. It is observed that by controlling the delay parameter in Eq. (\ref{field2}), the emission of the HHG cutoff energy can be temporally controlled systematically. The harmonic-cutoff energy emission time is found to scale as $t_{peak} \propto t_d^{5/2}$ with $t_d$ being the delay parameter [refer Eq. (\ref{field2})].

Furthermore, the harmonic yield in a couple of energy ranges and the harmonic cutoffs are also studied by varying $t_d$. An optimum delay parameter corresponding to the maximum harmonic cutoff energy and yield is also  observed. As a representative case to test the proof-of-concept, a different lattice parameter, say $d = 6$ a.u. in the potential $V(x)$ is also used and the results are presented in the Appendix \ref{appen1}.  The emission time of the cutoff energy with the delay parameter again showed the well-defined variation with different scaling parameter [$t_{peak}\propto t_d^{8/5}$]. This is expected as the electron dynamics for a given field profile depend solely on the band structure describing the periodic lattice, and hence the scaling parameter will have the imprint of the band structure of the periodic crystal under investigation. Further detailed analysis of the problem we reserve for the future. The detailed analysis of the chirp of the emitted harmonics for different lattice constant would also be an interesting study.

\section*{Acknowledgments} Authors would like to acknowledge the DST-SERB, Government of India, for funding the project CRG/2020/001020. 

\appendix
\renewcommand\thefigure{\thesection\arabic{figure}}    
\setcounter{figure}{0}    

\section{}
\label{appen1}

\begin{figure}[t]
\centering\includegraphics[width=1\columnwidth]{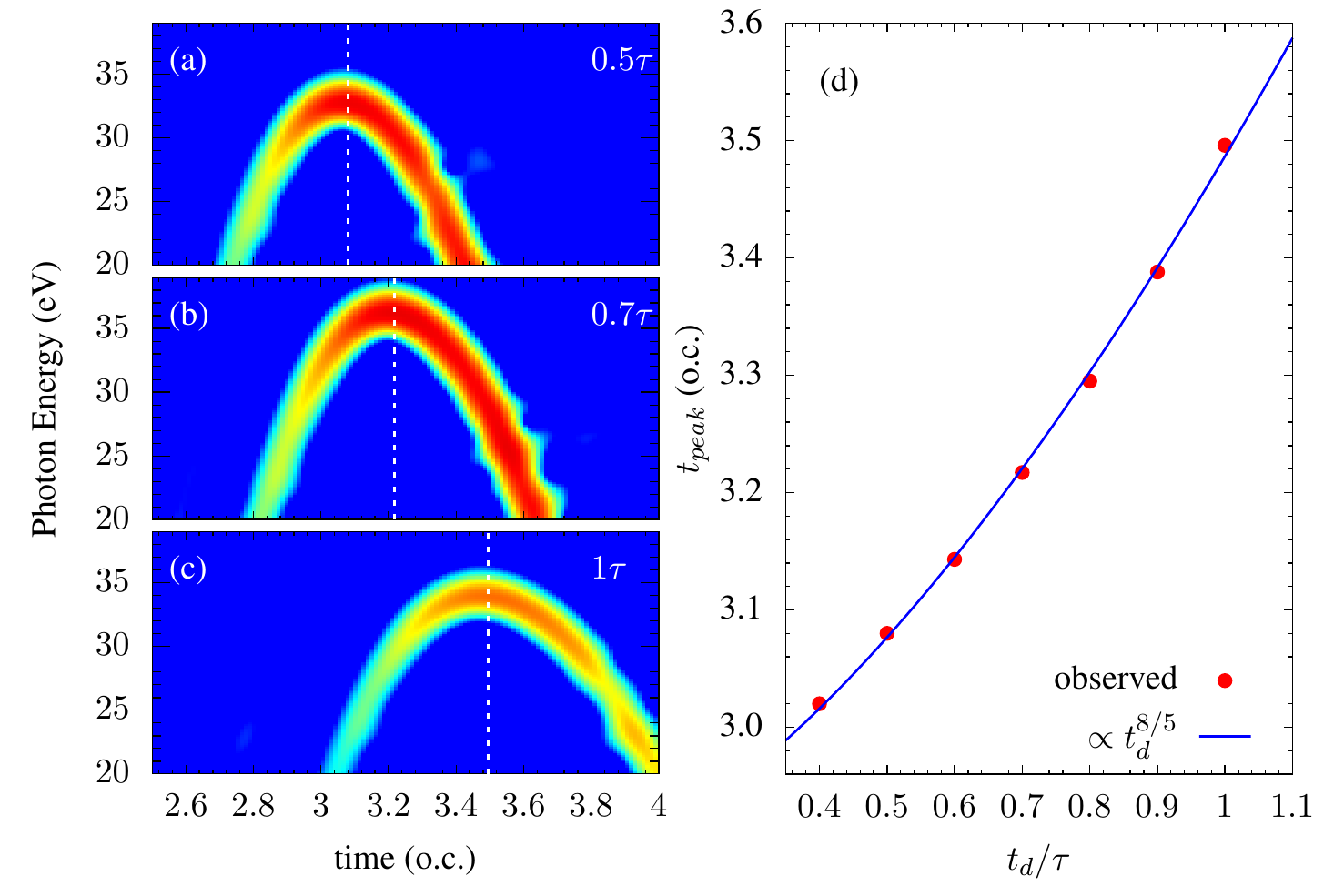}
\caption{The Gabor transform of the HHG spectra for the delay parameters $t_d = 0.5\tau$ (a), $0.7\tau$ (b), and $1\tau$ (c) are presented. All the other laser conditions are the same as in Fig. \ref{fig6}. A vertical dashed white line at $3.08\tau$ (a), $3.22\tau$ (b), and $3.49\tau$ (c) represents the time instants when the maximum energy harmonics are emitted. The variation of the time instant corresponding to the harmonic cutoff $t_{peak}$ with the delay parameter is presented in (d), wherein the red dots denote the observed values from the simulation and the solid line represents the $\propto t_d^{8/5}$ scaling.} 
\label{figA1}
\end{figure}
In order to further verify the capability of the sinc-shaped driver to temporally control the cutoff harmonic emission, we have carried out the simulations for the lattice parameter $d = 6$ a.u. using the same Mathieu-type potential \cite{PhysRev.87.807}. In Fig. \ref{figA1}(a), (b) and (c) we have presented the Gabor transform for HHG spectra emitted with $t_d = 0.5\tau$ (a), $0.7\tau$ (b) and $1\tau$ (c). The time at which maximum energy is emitted ($t_{peak}$) is plotted in Fig. \ref{figA1}(d) as function of the delay parameter, and it is observed that for this particular lattice parameter $t_{peak} \propto t_d^{8/5}$. 
 

%

\end{document}